\documentclass[twocolumn,preprintnumbers,superscriptaddress,amsmath,amssymb]{revtex4}

\usepackage{amsmath}
\usepackage{graphicx}
\usepackage{subfigure}
\usepackage{dcolumn}	
\usepackage{tabularray}
\usepackage{multirow}
\usepackage{hhline}
\usepackage{bm}
\usepackage{color}
\usepackage{booktabs, longtable}
\usepackage{CJK}
\usepackage{ltxtable}
\usepackage{rotating}
\usepackage[colorlinks, citecolor=red]{hyperref}
\usepackage{verbatim}
\usepackage{amsbsy}
\usepackage{amsfonts}
\usepackage{mathrsfs}
\usepackage{float}
\usepackage[english]{babel}

\begin{document}
\title{Quarkyonic matter with strangeness in an extended relativistic mean-field model}

\author{Wei Sun}
\affiliation{School of Physics, Zhengzhou University, Zhengzhou 450001, China}

\author{Cheng-Jun Xia}
\email[Corresponding author:~]{cjxia@yzu.edu.cn}
\affiliation{Center for Gravitation and Cosmology, College of Physical Science and Technology, Yangzhou University, Yangzhou 225009, China}

\author{Ting-Ting Sun}
\email[Corresponding author:~]{ttsunphy@zzu.edu.cn}
\affiliation{School of Physics, Zhengzhou University, Zhengzhou 450001, China}
	
\begin{abstract}
Quarkyonic matter is expected to play a key role for the transition from hadronic matter to quark matter in compact stars. Within the framework of the relativistic mean-field (RMF) model and equivparticle model with density-dependent quark masses, we construct the ``quark Fermi sea" with a ``baryon Fermi surface" to characterize the properties of the quarkyonic matter. In particular, we develop a comprehensive framework to account for the strangeness degrees of freedom, incorporating $\Lambda$, $\Xi$, and $\Sigma$ hyperons as well as strange quarks in a unified quarkyonic framework. Our calculations indicate that the inevitable emergence of hyperons softens the equations of state, leading to a reduction in the equilibrium sound velocity around $n_{\rm b}\approx 2n_0$, and consequently reducing the masses and radii of neutron stars. When the quark-hadron phase transition is taken into account, the equation of state at high densities exhibits additional softening consistent with current astronomical observational constraints. This softening leads to a maximum equilibrium sound velocity of $v_{eq}^{\rm max} \approx 0.6\,c$, which is close to the ultrarelativistic limit of $0.58\,c$.
\end{abstract}
	
\maketitle
	
\section{Introduction}
\label{Sec:Introduction}

Due to the asymptotic freedom and confinement properties, strongly interacting matter exhibits at least two distinct phases at zero temperature: hadronic matter (HM) and quark matter (QM). As baryon density increases, HM may undergo a deconfinement phase transition to QM. However, the precise mechanism underlying this transition remains unclear and multiple scenarios are possible~\cite{PPNP2023Voskresensky_130_104030,RPP2011Fukushima_74_14001}. 

One possibility is a quark-hadron mixed phase~\cite{PRL1993Heiselberg_70_1355,PR2001Glendenning_342_393,PLB2002Voskresensky_541_93,NPA2003Tatsumi_718_359,NPA2003Voskresensky_723_291,NPA2005Endo_749_333,PRD2007Maruyama_76_123015,PRC2008Peng_77_065807,PRC2014Yasutake_89_065803,PRD2019Xia_99_103017,PRC2019Maslov_100_025802,PRD2020Xia_102_023031}, in which the deconfinement transition between HM and QM is of first order~\cite{PRC2010Dexheimer_81_045201}. 
This transition can be described using the Gibbs and Maxwell construction~\cite{PRD1992Glendenning_46_1274,PR2001Glendenning_342_393,PRD2023Constantinos_107_074013}.
The Maxwell construction enforces charge neutrality locally and is suitable for systems with high interphase surface tension or for single-component systems. In contrast, the Gibbs construction imposes charge neutrality globally and applies when surface tension is negligible, making it more appropriate for multi-component charged systems. For intermediate values of the surface tension, the geometry of the phase boundary may change with density, resembling the structured ``pasta'' phases that appear in the crust-core transition region~\cite{PRL1993Heiselberg_70_1355}.

Another type of deconfinement phase transition is a smooth crossover from HM to QM, analogous to the transition observed at high temperature and vanishing chemical potential~\cite{PLB2014Borsányi_730_99,PRD2014Bazavov_90_094503}. At finite baryon number density, this hadron-quark crossover is often described phenomenologically using interpolation functions~\cite{PA1979Baym_96_131,PLB1980Celik_97_128}. Such approaches typically predict a stiffer equation of state (EOS), which is capable of supporting hybrid stars with masses up to approximately~$2M_{\odot}$~\cite{ApJ2013Masuda_764_12,PTEP2013Masuda_2013_073D01,PRD2015Zhao_92_054012,PRD2015Kojo_91_045003,EPJA2016Masuda_52_65,PRC2016Whittenbury_93_035807,PRD2017Li_95_056018,PRD2018Li_98_083013,PRD2018Zhan_97_023018,ApJ2019Baym_885_42}. To clarify the microscopic dynamics underlying this crossover regime, quarkyonic matter phase has been proposed as a key mechanism for the transition from hadronic to quark matter~\cite{ApJ2016Fukushima_817_180,NPA2009McLerran_830_709c,NPA2007McLerran_796_83}. This concept addresses the prominent role of many-body interactions among baryons at supra-saturation densities~\cite{PRL2010Hebeler_105_161102}. These enhanced interactions are attributed to a greater number of exchanged quarks as density increases ~\cite{ApJ2016Fukushima_817_180}. When density exceeds a critical threshold, the distinct boundaries between individual baryons gradually diminish. This allows quarks to delocalize and move more freely among baryons, leading to the formation of the quarkyonic phase~\cite{NPA2009McLerran_830_709c}. 
To describe the quarkyonic matter, Cao~\emph{et al.} developed a unified field theory by combining the Walecka model with the quark-meson model, which can treat baryons, quarks, and mesons on equal footing, enabling a consistent description of both chiral symmetry breaking and restoration processes within the quarkyonic phase~\cite{JHEP2020Cao_10_168,PRD2022Cao_105_114020}. Besides, Zhao~\emph{et al.} proposed a quarkyonic model including protons and leptons in chemical and beta equilibrium that satisfies the experimental and observational constraints on neutron star structure~\cite{PRD2020Zhao_102_023021}.
Recently, we also developed a unified framework to model quarkyonic matter by combining the relativistic mean field (RMF) model and the equivparticle model with density-dependent quark masses, which can give a coherent investigation of nuclear matter, quark matter, and quarkyonic matter in a unified manner~\cite{PRD2023Xia_108_054013}. Our results show that the onset of the quarkyonic phase leads to a softening of the equation of states (EOSs) and can accommodate various experimental and astrophysical constraints generally.

The study of strange quark matter is of significant importance for exploring new forms of matter, understanding the mechanism of Quantum Chromodynamics (QCD) phase transitions, and investigating the nature of strong interactions. A detailed investigation of its properties is essential to understand the deconfinement phase transition~\cite{PRC2008Peng_77_065807,PRD2013Logoteta_88_063001,PRC2014Orsaria_89_015806}, the restoration of chiral symmetry~\cite{PLB2002Peng_548_189,CPL2014Xia_31_041101,PLB2022Jin_829_137121}, the properties of high-temperature and high-density matter produced in relativistic heavy-ion collisions~\cite{NPA2013Wiedemann_904_3c}, as well as the structure of compact stars~\cite{PPNP2005Weber_54_193,AA2007Bombaci_462_1017,PRD2018Xia_98_034031,PRD2019Sun_99_023004}. The appearance of hyperons will soften the EOSs, affecting the prediction of neutron star mass and radius~\cite{PRC2014Lopes_89_025805,PLB2015Katayama_747_43,PRD2019Sun_99_023004,PRD2025Ahmad_112_023013}. However, a soft EOS will result in compact stars with too small masses that cannot reach two solar mass $2M_{\odot}$ as observed in pulsars PSR J1614-2230 ($1.928 \pm 0.017 M_{\odot}$)~\cite{Nature2010Demorest_467_1081,ApJ2016Fonseca_832_167} and PSR J0348$+$ 0432~($2.01 \pm 0.04 M_{\odot}$)~\cite{Science2013John_340_1233232}, leading to the ``Hyperon Puzzle''~\cite{AIP2015Vidaña_1645_79}. Substantial theoretical efforts have been devoted to resolve this puzzle, including approaches that modify hyperonic interactions, incorporate additional repulsive mechanisms, or introduce new degrees of freedom~\cite{PRC2012Weissenborn_85_065802,PRL2015Lonardoni_114_092301,EPJA2016Masuda_52_65,PRD2015Zhao_92_054012,ApJ2016Fukushima_817_180,CPC2018Sun_42_025101,PLB2025Sun_865_139460}. Recently, a promising direction has emerged through models that explicitly account for the quark substructure of baryons. In particular, the ideal dual quarkyonic (IdylliQ) model proposed by Fujimoto~\emph{et al.}~\cite{PRL2024Fujimoto_132_112701,EPJ2025Fujimoto_316_07007} suggests that the internal quark dynamics can delay the onset of hyperons and thereby mitigate the softening of the EOS, offering a coherent pathway to reconcile hyperon formation with the observed massive neutron stars.

In this work, we characterize the properties of quarkyonic matter by constructing a “quark Fermi sea” and a “baryon Fermi surface” within a density-dependent quark mass equivparticle model. As an extension of our previous work~\cite{PRD2023Xia_108_054013}, we further introduce strangeness degrees of freedom, incorporating $\Lambda$, $\Xi$, and $\Sigma$ hyperons together with strange quarks into a unified quarkyonic framework. For the description of baryonic matter, we adopt the meson-exchange RMF model~\cite{BookMeng}, which has been well established in describing both finite nuclei~\cite{RPP1989Reinhard_52_439,PPNP1996Ring_37_193,PPNP2006Meng_57_470,JPG2015Meng_42_093101,NPA1999Typel_656_331,PRC2011Lu_84_014328,NPA2012Belvedere_883_1,PRC2019Sun_99_034310,PLB2023Sun_847_138320,PLB2024Sun_854_138721,PRC2024Sun_109_014323} and nuclear matter~\cite{Glendenning2000,PRC2004Ban_69_045805,PPNP2007Weber_59_94,PRC2012Sun_86_014305,PRC2014Wang_90_055801,PRC2015Fedoseew_91_034307,PRC2025Xia_112_064904}. To describe the quarkyonic matter, the RMF approach is combined with an equivparticle model with density-dependent quark masses~\cite{JPSCP2018Xia_20_011010,PRD2023Xia_108_054013,PRL2019McLerran_122_122701}. The coupling between quark and baryon sectors is implemented through density-dependent effective baryon masses, and the microscopic properties of quarkyonic matter are obtained via energy minimization.

This paper is organized as follows: In Sec.~\ref{Sec:Theory}, we present the theoretical framework for the meson-exchange RMF model and equivparticle model. In Sec.~\ref{Sec:Numerical}, numerical details including the baryon-baryon interactions and parameter sets of the density-dependent mass scalings for baryons and quarks are given. After the properties of dense matter and their implications for compact star structures analyzed in Sec.~\ref{Sec:Results}, we summarize in Sec.~\ref{Sec:Summary} finally. The natural units $\hbar = c = 1$ is used throughout the text.

\section{Theoretical framework}
\label{Sec:Theory}

The Lagrangian density of the extended RMF model can be decomposed into three parts as the following,
\begin{equation}
 \mathcal{L}=\mathcal{L}^{\rm B}+\mathcal{L}^{\rm Q}+\mathcal{L}^{\rm L}, 
\end{equation}
where $\mathcal{L}^{\rm B}, \mathcal{L}^{\rm Q}$, and $\mathcal{L}^{\rm L}$ represent the Lagrangian densities for baryons, quarks, and leptons, respectively,
\begin{subequations}
\begin{eqnarray}
 \mathcal{L}^{\rm B}&=&{\sum_{b=n,p,\Lambda,\Xi,\Sigma}}\bar{\Psi}_{b}\{i\gamma^{\mu}\partial_{\mu}-{ m_b}(n_{\rm b}^{\rm Q})-g_{\sigma b}(n_{\rm b}^{\rm B})\sigma\nonumber \\
 &&-{g_{\omega b}}(n_{\rm b}^{\rm B})\gamma^{\mu}\omega_{\mu}-g_{\rho b}(n_{\rm b}^{\rm B})\gamma^{\mu}
 \bm{ \tau}_{b}\cdot\bm{\rho}_{\mu} \}\Psi_{b}\nonumber\\
 &&-\displaystyle\frac{1}{2}m_{\sigma}^{2}\sigma^{2}+\displaystyle\frac{1}{2}m_{\omega}^{2}\omega_{\mu}\omega^{\mu}+\displaystyle\frac{1}{2}m_{\rho}^{2}\bm{\rho}_{\mu}\cdot\bm{\rho}^{\mu},~\label{Eq:Lagbaryons}\\
 \mathcal{L}^{\rm Q}&=&{\sum_{q=u,d,s}}\bar{\Psi}_{q}[i\gamma^{\mu}\partial_{\mu}-m_{q} (n_{\rm b})]\Psi_{q},~\label{Eq:Lagquark}\\	
 \mathcal{L}^{\rm L}&=&\sum_{l=e,\mu}\bar{\Psi}_{l}[i\gamma^{\mu}\partial_{\mu}-m_{l}]\Psi_{l}.~\label{Eq:Lagleptons}
\end{eqnarray}%
\label{Eq:Lagrangian}% 
\end{subequations}%
Here $\Psi_{i}~(i=b, q, l)$ denotes the Dirac spinor for different fermions of mass $m_{i}$, i.e., baryons comprising nucleons ($n, p$) and hyperons ($\Lambda$, $\Xi$, $\Sigma$), $u, d, s$ quarks, and leptons including electrons ($e$) and muons ($\mu$). The isospin of baryons is represented by $\bm\tau_b$. 
The baryon-baryon interactions are described by the exchange of the isoscalar-scalar $\sigma$ meson, the isoscalar-vector $\omega_{\mu}$ meson, and the isovector-vector $\bm\rho_{\mu}$ meson with $m_\xi$~($\xi=\sigma$, $\omega$, $\rho$) and $g_{\xi b}$ being the corresponding mass and the coupling constant with baryons, respectively. In a system with time-reversal symmetry, the spacelike components of the vector fields $\omega_{\mu}$ and $\bm{\rho}_{\mu}$ vanish, leaving only the time components $\omega_{0}$ and $\bm{\rho}_{0}$. Furthermore, charge conservation guarantees only the third component $\rho_{0,3}$ in the isospin space survives. It is noted that, the $\sigma$, $\omega_{0}$, and $\rho_{0,3}$ fields are spatially constant for uniform dense matter, thus both their spatial and time derivatives vanish. In what follows, the notations $\omega$ and $\rho_3$ are adopted for simplicity to represent these fields. 

In the quarkyonic phase, baryons and quarks coexist within the same volume. The baryon number densities $n_{\rm b}$ are contributed from both baryons ($n_{\rm b}^{\rm B}$) and quarks ($n_{\rm b}^{\rm Q}$),
\begin{subequations}
\begin{align}
 &n_{\rm b}=n_{\rm b}^{\rm B}+n_{\rm b}^{\rm Q},\\ 
 &n_{\rm b}^{\rm B}=n_n+n_p+n_{\Lambda}+n_{\Xi}+n_{\Sigma},\\ 
 &n_{\rm b}^{\rm Q}={(n_u+n_d+n_{s})}/3.
\end{align}	
\label{Eq:nb}
\end{subequations}	

In the Lagrangian density~(\ref{Eq:Lagrangian}), masses both for baryons and quarks are density-dependent. For baryons, a phenomenological mass scaling~\cite{PRD2023Xia_108_054013} is employed to account for the effects of Pauli blocking and the interactions between quarks and baryons, i.e.,  
\begin{equation}
m_b(n_{\rm b}^{\rm Q})=m_{0b}+Bn_{\rm b}^{\rm Q},
\label{Eq:baryon mass}
\end{equation}
where $m_{0b}$ represents the baryon mass in vacuum, and $B$ is the interaction strength.

For quarks, they are treated as quasifree particles with density-dependent masses in the framework of equivparticle models~\cite{PRC1999Peng_61_015201, PLB2002Peng_548_189,PRC2005Wen_72_015204,PRD2014Xia_89_105027}. By incorporating linear confinement and leading-order perturbative interactions, the quark mass scaling is taken as follows~\cite{PRD2014Xia_89_105027},
\begin{equation}
m_q(n_{\rm b}) = m_{0q} + \frac{D}{\sqrt[3]{n_{\rm b}}} + C\sqrt[3]{n_{\rm b}},
\label{Eq:quark mass}
\end{equation}
where $m_{0u}=2.3$~MeV, $m_{0d}=4.8$~MeV, and $m_{0s}=95$~MeV are the current quark masses~\cite{CPC2014Olive_38_090001}. The parameter $D$ characterizes the confinement strength, which is closely related to the chiral restoration density, the string tension, and the sum of the vacuum chiral condensates. The parameter $C$ represents the perturbative strength associated with the strong coupling constant.
Although the values of $D$ and $C$ are not yet determined explicitly due to uncertainties in the relevant quantities, estimates suggest that $\sqrt{D}$ lies approximately in the range $147$-$270$~MeV~\cite{PRC2005Wen_72_015204} while $C \lesssim 1.2$~\cite{PRD2014Xia_89_105027}.
	
In accordance with the Typel-Wolter ansatz~\cite{NPA1999Typel_656_331}, density-dependent nucleon-meson coupling constants $g_{\xi b}(n_{\rm b}^{\rm B})$ are employed in the Lagrangian~(\ref{Eq:Lagrangian}). For the $\sigma$ and $\omega$ mesons, the coupling constants are determined by,  
\begin{equation}
 g_{\xi b}({n_{\rm b}^{\rm B}})=g_{\xi b}(n_0)a_\xi\frac{1+b_\xi(x+d_\xi)^2}{1+c_\xi(x+e_\xi)^2},
 \label{Eq:gsiggome}
\end{equation}
where $g_{\xi b}(n_0)$ is the value at the saturation density $n_0\approx 0.16$~fm$^{-3}$ of nuclear matter or for the finite nuclei, $ x \equiv {n_{\rm b}^{\rm B}}/{n_0}$, and $a_\xi$, $b_\xi$, $c_\xi$, $d_\xi, $ and $e_\xi$ represent five adjustable coefficients which describe the density-dependent coupling constants. Meanwhile, for the coupling constant of $\rho$ meson, an alternative formula is employed,
\begin{equation}
g_{\xi b}({n_{\rm b}^{\rm B}})=g_{\xi b}(n_0)\exp[-a_\xi(x+b_\xi)].
 \label{Eq:rho}
\end{equation}
		
Based on the Lagrangian density in Eq.(\ref{Eq:Lagrangian}), the meson fields are determined by,  
\begin{subequations}
\begin{align}
 m_{\sigma}^{2}\sigma&=-{\sum_{b=n,p,\Lambda,\Xi,\Sigma}}g_{\sigma b}(n_{\rm b}^{\rm B})n_{b}^{s},\\
 m_{\omega}^{2}\omega&={\sum_{b=n,p,\Lambda,\Xi,\Sigma}}g_{\omega b}(n_{\rm b}^{\rm B})n_{b}, \\
 m_{\rho}^{2}\rho_{3}&=\sum_{b=n,p,\Xi,\Sigma}g_{\rho b}(n_{\rm b}^{\rm B})\tau_{b,3}n_{b}, 
\end{align}%
\label{Eq:meson}%
\end{subequations}%
where the source currents $n_b$ and $n_b^s$ are respectively the baryon number density and scalar density. Employing the no-sea approximation, these densities for fermions $(i=b, q, l)$ are calculated by,  
\begin{subequations}
\begin{align}
 &n_{i}=\langle\bar{\Psi}_{i}\gamma^{0}\Psi_{i}\rangle=\frac{g_{i}\nu_{i}^{3}}{6\pi^{2}},\\
 &n_{i}^{s}=\langle\bar{\Psi}_{i}\Psi_{i}\rangle=\frac{g_{i}(m_{i}^{*})^{3}}{4\pi^{2}}\biggl[x_{i}\sqrt{x_{i}^{2}+1}-{\rm arcsh}(x_{i})\biggr],
\end{align}%
\label{Eq:density}%
\end{subequations}%
where $g_{i}$ is the degeneracy factor, and for baryons, quarks, and leptons, we take  $g_{n,p,\Lambda,\Xi,\Sigma}=2$, $g_{u,d,s}=6$, and $g_{e,\mu}=2$, respectively. Here, we defined $x_i\equiv \nu_i/m_i^*$, with $\nu_i$ being the Fermi momentum and $m_i^*$ the effective mass for baryons and quarks,
\begin{subequations}
\begin{align}
 &m_b^* = m_b(n_{\rm b}^{\rm Q})+g_{\sigma b}\sigma,\\
 &m_q^*=m_q(n_{\rm b}),
\end{align}%
\label{Eq:mass}%
\end{subequations}%
with baryon mass scaling $m_b(n_{\rm b}^{\rm Q})$ in Eq.~(\ref{Eq:baryon mass}) and quark mass scaling $m_q(n_{\rm b})$ in Eq.~(\ref{Eq:quark mass}). Meanwhile, the masses of leptons remain constant, i.e., $m_e^*=0.511$~MeV and $ m_\mu^*=105.66$~MeV~\cite{CPC2014Olive_38_090001}. 

The single-particle energies of fermions at fixed momentum $p$ are given as,
\begin{subequations}
\begin{align}
 &\epsilon_{b}(p)=g_{\omega b}\omega+g_{\rho b}\tau_{b,3}\rho_{3}+\Sigma_{b}^{\rm R}+\sqrt{p^{2}+(m_{b}^{*})^{2}}, \\
 &\epsilon_{q}(p)=\Sigma_{q}^{\rm R}+\sqrt{p^{2}+(m_{q}^{*})^{2}}, \\
 &\epsilon_{l}(p)=\sqrt{p^{2}+(m_{l}^{*})^{2}},
\end{align}
\label{Eq:Esp}
\end{subequations}
where the ``rearrangement'' terms are defined as  
\begin{subequations}
\begin{align}
 \Sigma_{b}^{\rm R}&=\sum_{b}
 \left(\frac{{\rm d}g_{\sigma b}}{{\rm d}n_{\rm b}^{\rm B}}\sigma n_{b}^{s}+\frac{{\rm d}g_{\omega b}}{{\rm d}n_{\rm b}^{\rm B}}\omega n_{b}+\frac{{\rm d}g_{\rho b}}{{\rm d}n_{\rm b}^{\rm B}}\tau_{b,3}\rho_{3}n_{b}\right)\notag\\ 
 &+{\sum_{q}}\frac{{\rm d}m_q}{{\rm d}n_{\rm b}}n_q^s,\\ 
\Sigma_{q}^{\rm R}&=\frac{1}{3}{\sum_{i=b,q}}\frac{{\rm d}m_{i}}{{\rm d}n_{\rm b}}n_{i}^{s}. 	
\end{align}
\label{Eq:rearrangement}
\end{subequations}

In the quarkyonic phase, an inner Fermi sea of quarks is surrounded by an outer shell of baryons. 
At their interface, a single baryon with zero momentum ($p_{b}=0$) is matched to three quarks, enforcing that its energy equals the sum of the corresponding quark energies at that point, i.e.,
\begin{subequations}
\begin{align}
 &\epsilon_{u}(\nu_{u})+2\epsilon_{d}(\nu_{d})=\epsilon_{n}(0),\\
 &2\epsilon_{u}(\nu_{u})+\epsilon_{d}(\nu_{d})=\epsilon_{p}(0),\\	
 &\epsilon_{u}(\nu_{u})+\epsilon_{d}(\nu_{d})+\epsilon_{s}(\nu_{s})=\epsilon_{\Lambda}(0). %
\end{align}%
\label{Eq:ChemicalEqu}%
\end{subequations}%
It is noted that $\nu_u$, $\nu_d$, and $\nu_s$ here correspond to the maximum momenta for $u$, $d$, and $s$ quarks, respectively, rather than Fermi momenta.

Based on the single particle energy at the Fermi momentum $\nu_i~(i=b,l)$, the chemical potentials for baryons and leptons are determined by $\mu_b=\epsilon_b(\nu_b)$ and $\mu_l=\epsilon_l(\nu_l)$, respectively. Formally, an effective chemical potential $\mu_q$ for quarks can also be defined as $\mu_q=\epsilon_q(\nu_q)$. However, this quantity does not represent a true chemical potential, as $\nu_q$ does not correspond to the Fermi surface in the quarkyonic phase.

For neutron star matter, the charge neutrality condition should be fulfilled, 
\begin{equation}
 \sum_iq_in_i=0,
 \label{Eq:charge}
\end{equation}
where $q_{i}$ represent the charge numbers of each particle type. In order to reach the lowest energy state, the particles will undergo evolution through weak interactions until they satisfy the $\beta$ equilibrium condition, 
\begin{equation}
\mu_{b}=\mu_n-q_{b}\mu_e, \quad\mu_\mu=\mu_e.
\label{Eq:betaEqu}
\end{equation}

Finally, the energy density $\varepsilon$ can be determined as follows,
\begin{equation}
\varepsilon=\sum_{i}\tau_i(\nu_i,m_i^*)+\sum_{\xi=\sigma,\omega,\rho}\frac{1}{2}m_\xi^2\xi^2,
 \label{Eq:energy}
\end{equation}
where $\tau_i$ is the kinetic energy density for different fermions, 
\begin{align}
\tau_i(\nu_i,m_i^*)&=\int_{0}^{\nu_{i}}\frac{g_{i}p^{2}}{2\pi^{2}}\sqrt{p^{2}+(m_{i}^{*})^{2}}dp \\
 &=\frac{g_i(m_i^*)^4}{16\pi^2}\left[x_i(2x_i^2+1)\sqrt{x_i^2+1}-{\rm arcsh}(x_i)\right].\nonumber 
\label{Eq:kineticE}%
\end{align}%
The pressure $P$ is subsequently obtained as follows,
\begin{equation}
 P=\sum_i\mu_in_i-\varepsilon.
 \label{Eq:pressure}
\end{equation}

In summary, the extended RMF theory with the strangeness degree of freedom can describe the nuclear matter~($n_{\rm b}=n_{\rm b}^{\rm B}$), quark matter~($n_{\rm b}=n_{\rm b}^{\rm Q}$), and quarkyonic matter~($n_{\rm b}=n_{\rm b}^{\rm B}+n_{\rm b}^{\rm Q}$) on same footing. At a given total baryon number density $n_{\rm b}$, the densities in Eq.(\ref{Eq:nb}), the mean fields in Eq.(\ref{Eq:meson}), and the single-particle energies in Eq.(\ref{Eq:Esp}) are obtained by iteratively solving Eqs.(\ref{Eq:baryon mass})–(\ref{Eq:ChemicalEqu}) with the single-particle energy relationship in Eq.(\ref{Eq:ChemicalEqu}), charge neutrality condition in Eq.(\ref{Eq:charge}), and the $\beta$ equilibrium condition in Eq.(\ref{Eq:betaEqu}). After achieving convergence, the energy density $\varepsilon$ and pressure $P$ can be calculated using Eqs.(\ref{Eq:energy}) and (\ref{Eq:pressure}), with which the EOSs of neutron star matter can be finally obtained.

Based on the obtained EOSs, the corresponding structures of compact stars are determined by solving the Tolman-Oppenheimer-Volkov (TOV) equation,
\begin{equation}
 \frac{{\rm d}P}{{\rm d}r}=-\frac{GM\varepsilon}{r^2}\frac{(1+P/\varepsilon)(1+4\pi r^3P/M)}{1-2GM/r},~\label{TOV}
\end{equation}
with the subsidiary condition
\begin{equation}
 \frac{{\rm d}M}{{\rm d}r}=4\pi\varepsilon r^2.
\end{equation}
 
The dimensionless tidal deformability is calculated by
\begin{equation}
 \Lambda=\frac{2k_2}{3}\left(\frac{R}{GM}\right)^5,
\end{equation}
where the second Love number $k_2 $ is evaluated by introducing perturbations to the metric~\cite{PRD2009Damour_80_084035,PRD2010Hinderer_81_123016,PRD2010Postnikov_82_024016}.

\section{Numerical Details}
\label{Sec:Numerical}
%------------------------Table 1-----------------------	     
\begin{table*}[t!]
	\centering
	\caption{Properties of hyperons ($\Lambda$, $\Xi^{0,-}$, $\Sigma^{+,0,-}$) including the quark components, mass in vacuum $m_0$~(MeV), third component of isospin $\tau_{3}$, total angular momentum and parity $J^{p}$, and electric charge $q$. Also listed are the relative ratios $\alpha_{\xi} = g_{\xi Y}/g_{\xi N}$ between the hyperon-meson couplings $g_{\xi Y}$ and nucleon-meson couplings $g_{\xi N}$ with different mesons $\xi = \sigma, \omega, \rho$.}
	\begin{tabular}{lccrcrccccc} 
		\hline\hline
		             & \multirow{2}{*}{quarks} & \multirow{2}{*}{$m_0$} & \multirow{2}{*}{$\tau_3$} & \multirow{2}{*}{$J^P$} & \multirow{2}{*}{$q$} & \multicolumn{3}{c}{$\alpha_{\sigma}$} & \multirow{2}{*}{$\alpha_{\omega}$} & \multirow{2}{*}{$\alpha_{\rho}$} \\
		\cline{7-9}
		             & & & & & & TW99 & PKDD & DD-ME2 & & \\
		\hline
		$\Lambda$    & uds   & 1115.6  & 0   & $(1/2)^{+}$ & 0    & 0.887 & 0.888 & 0.887 & 1 & 0 \\
		$\Xi^{0}$    & uss   & 1314.9  & +1  & $(1/2)^{+}$ & 0    & 0.854 & 0.855 & 0.854 & 1 & 1 \\
		$\Xi^{-}$    & dss   & 1321.3  & -1  & $(1/2)^{+}$ & $-e$ & 0.854 & 0.855 & 0.854 & 1 & 1 \\
		$\Sigma^{+}$ & uus   & 1189.4  & +2  & $(1/2)^{+}$ & $+e$ & 0.744 & 0.740 & 0.739 & 1 & 1 \\
		$\Sigma^{0}$ & uds   & 1192.5  & 0   & $(1/2)^{+}$ & 0    & 0.744 & 0.740 & 0.739 & 1 & 1 \\
		$\Sigma^{-}$ & dds   & 1197.4  & -2  & $(1/2)^{+}$ & $-e$ & 0.744 & 0.740 & 0.739 & 1 & 1 \\
		\hline\hline
	\end{tabular}
	\label{Table1}
\end{table*}

As extension of our previous work~\cite{PRD2023Xia_108_054013}, this work incorporates hyperons ($Y=\Lambda$, $\Xi^{0,-}$, $\Sigma^{+,0,-}$) and strange quarks ($s$) to describe the properties of the quarkyonic matter. In Table~\ref{Table1}, we list the properties of different hyperons including the quark components, mass in vacuum $m_0$, third component of isospin $\tau_{3}$, total angular momentum and parity $J^{p}$, and electric charge $q$. 

%------------------------Table 2-----------------------	   
\begin{table}
\centering
\caption{Saturation properties of nuclear matter including the saturation density $n_{0}$, the binding energy per nucleon $B$, incompressibility $K$, skewness coefficient $Q$, symmetry energy $J$, slope $L$, and curvature parameter $K_{\rm sym}$ of nuclear symmetry energy predicted by three different density-dependent covariant density functionals TW99~\cite{NPA1999Typel_656_331}, PKDD~\cite{PRC2004Long_69_034319}, and DD-ME2~\cite{PRC2005Lalazissis_71_024312}.}
\begin{tabular}{llllllll} 
\hline\hline
      & $n_{0}$   & $B$~~~~~~&$K$~~~~~~&$Q$~~~~~~&$J$~~~~~~&$L~~~~~~$ &$K_{\rm sym}$ \\ \cline{2-8}
	  & fm$^{-3}$ & MeV      & MeV     & MeV     & MeV     & MeV      & MeV   \\\hline
TW99  & $0.153$   & $-16.24$ & $240.2$ &$-540$   & $32.8$  & $55.3$   & $-125$ \\			
PKDD  & $0.150$   & $-16.27$ & $262.2$ &$-119$   & $36.8$  & $90.2$   & $-81$   \\			
DD-ME2& $0.152$   & $-16.13$ & $250.8$ &~~$477$  & $32.3$  & $51.2$   & $-87$   \\\hline\hline
\end{tabular}
\label{Table2}
\end{table}
	
For baryonic matter described by the Lagrangian density in Eq.~(\ref{Eq:Lagbaryons}), we adopt three distinct density-dependent covariant density functionals for the nucleon-nucleon interactions, TW99~\cite{NPA1999Typel_656_331}, PKDD~\cite{PRC2004Long_69_034319}, and DD-ME2~\cite{PRC2005Lalazissis_71_024312} following our previous work~\cite{PRD2023Xia_108_054013}. Based on these interactions, Table~\ref{Table2} lists the properties of nuclear matter near the saturation density ($n_0 \approx 0.16~{\rm fm}^{-3}$). Several key parameters are well constrained: the binding energy $B \approx -16$~MeV, incompressibility $K = 240 \pm 20$~MeV~\cite{EPJA2006Shlomo_30_23}, symmetry energy $J= 31.7 \pm 3.2$~MeV, and slope parameter $L=58.7 \pm 28.1$~MeV~\cite{PLB2013Li__727_276,RMP2017Oertel_89_015007}. Among the chosen functionals, TW99 yields saturation properties in good agreement with these constraints.
In contrast, PKDD predicts a larger symmetry energy (larger $J$ and $L$) while the energy per baryon for symmetric nuclear matter at suprasaturation densities is significantly increased (larger $K$ and $Q$) if DD-ME2 is adopted. For more discussions, see Ref.~\cite{PRD2023Xia_108_054013}.

The nucleon-hyperon interaction are fixed by reproducing the available experimental data or empirical values. Many works for finite hypernuclei in the framework of RMF theory have been performed~\cite{BookMeng,PRC2017Sun_96_044312,JPG2017Lu_44_125104,SCPMA2021Chen_64_282011,CPC2022Sun_46_074106} where the nucleon-hyperon interaction have been discussed widely~\cite{PRC2016Sun_94_064319,PRC2017Ren_95_054318,PRC2018Liu_98_024316,CPC2018Sun_42_025101,PRC2022Tanimura_105_044324,PRC2025Ding_111_014301}.
For the $\Lambda$ hyperons, which are neutral and isospin scalars, only couplings with the $\sigma$ and $\omega$ mesons are considered.
According to our previous studies~\cite{CPC2018Sun_42_025101,PRD2019Sun_99_023004}, the mass of PSR J0348+0432 can only be achieved with large values of $g_{\omega\Lambda}$. Therefore, in this work we adopt $g_{\omega \Lambda} = g_{\omega N}$, which yields $g_{\sigma \Lambda} = 0.888g_{\sigma N}$ under the constraint of the mean-field potential depth $V_{\Lambda} = -29.786$~MeV for $\Lambda$ hyperons in symmetric nuclear matter at saturation density~\cite{CPC2018Sun_42_025101}.
For the $\Xi$ and $\Sigma$ hyperons, which are isospin vectors and may charged, couplings with the $\sigma$, $\omega$, $\rho$ mesons and photons are all included. By taking $g_{\omega \Xi}=g_{\omega \Sigma}=g_{\omega N}$, the couplings with the $\sigma$-meson are determined as $g_{\sigma \Xi}=0.855g_{\sigma N}$ and $g_{\sigma \Sigma}=0.740g_{\sigma N}$, under the constraints of the mean-field potential depths $V_{\Xi}=-16.276$~MeV~\cite{PRC2018Liu_98_024316} for $\Xi$ hyperons and $V_{\Sigma}=30$~MeV~\cite{PRC2000Avraham_62_034311,PLB2015Katayama_747_43,PRC2018Liu_98_024316,PRD2025Ahmad_112_023013} for $\Sigma$ hyperons, respectively.
For the couplings with $\rho$-mesons, we set $g_{\rho \Xi}=g_{\rho \Sigma}=g_{\sigma N}$, consistent with their isospin symmetry properties~\cite{PRC2016Sun_94_064319,PRC2018Liu_98_024316}. In Table~\ref{Table1}, the hyperon-meson couplings $g_{\xi Y}$~$(\xi = \sigma, \omega, \rho)$ are summarized by giving the relative ratios $\alpha_{\xi} = g_{\xi Y}/g_{\xi N}$ with respect to the nucleon-meson couplings $g_{\xi N}$.

%------------------------Table 3-----------------------	
\begin{table}[t!]
\centering
\caption{Parameter sets $(B, C, \sqrt{D})$ adopted for the baryon and quark mass scalings in Eqs.~(\ref{Eq:baryon mass}) and~(\ref{Eq:quark mass}). Additionally, the predicted radii $R_{1.4}$ and tidal deformability $\Lambda_{1.4}$ of 1.4-solar-mass compact stars, as well as the maximum mass $M_{\rm TOV}$ and the maximum equilibrium sound velocity $v_{eq}^{\rm max}$ of quarkyonic matter are provided.}
\begin{tabular}{lccccccc} 
\hline\hline
\multirow{2}{*}{}& $B$   & \multicolumn{2}{l}{$C$~~~$\sqrt{D}$~~~} & \multicolumn{3}{l}{$R_{1.4}$~~~$\Lambda_{1.4}$~~~$M_{\rm TOV}$} & $v_{eq}^{\rm max}$ \\ \cline{2-4} \cline{5-8}
	             & MeV/fm$^{3}$  &  &MeV       & km         &                 &$M_{\odot}$         & $c$  \\\hline
TW99             &$0$      & $0$    &~~~$0$    & $12.31$    &~~$514$          &~~$1.97$            & $0.89$  \\
		         &$0$      & $0.7$  &~~$180$   & $12.31$    &~~$514$          &~~$1.88$            & $0.65$  \\
		         &$300$    & $0.7$  &~~$180$   & $12.31$    &~~$514$          &~~$1.95$            & $0.71$  \\
		         &$300$    & $0.2$  &~~$180$   & $12.24$    &~~$421$          &~~$1.80$            & $0.66$  \\
		         &$300$    & $0.7$  &~~$230$   & $12.31$    &~~$514$          &~~$1.97$            & $0.83$  \\\hline
PKDD             &$0$      & $0$    &~~~$0$    & $13.67$    &~~$872$          &~~$2.24$            & $0.91$  \\
	             &$150$    & $0.7$  &~~$150$   & $12.84$    &~~$639$          &~~$2.01$            & $0.64$  \\
		         &$350$    & $0.7$  &~~$150$   & $13.28$    &~~$742$          &~~$2.10$            & $0.65$  \\
		         &$350$    & $0.5$  &~~$150$   & $12.51$    &~~$568$          &~~$2.00$            & $0.65$  \\
		         &$150$    & $0.7$  &~~$180$   & $13.67$    &~~$872$          &~~$2.15$            & $0.64$  \\\hline
DD-ME2           &$0$      & $0$    &~~~$0$    & $13.23$    &~~$748$          &~~$2.42$            & $0.92$  \\
		         &$100$    & $0.5$  &~~$160$   & $12.77$    &~~$580$          &~~$2.00$            & $0.62$  \\
		         &$350$    & $0.5$  &~~$160$   & $13.16$    &~~$714$          &~~$2.14$            & $0.63$  \\
		         &$350$    & $0.3$  &~~$160$   & $12.58$    &~~$556$          &~~$2.01$            & $0.63$  \\
	             &$100$    & $0.5$  &~~$0.5$   & $13.23$    &~~$748$          &~~$2.15$            & $0.61$  \\
\hline\hline
\end{tabular}
\label{Table3}
\end{table}	

Based on the aforementioned density functionals, we further consider the potential formation of quarkyonic matter by explicitly incorporating quasifree quarks. The adopted parameter sets $(B, C, \sqrt{D})$ for the baryon and quark mass scalings in Eqs.~(\ref{Eq:baryon mass}) and~(\ref{Eq:quark mass}) are listed in Table~\ref{Table3}, where $B$ is in MeV/fm$^{3}$, $C$ is dimensionless, and $\sqrt{D}$ is in MeV. 

%------------------------------Fig 1----------------------------------------

\begin{figure}[t!]
	\centering
	\includegraphics[width=0.95\linewidth]{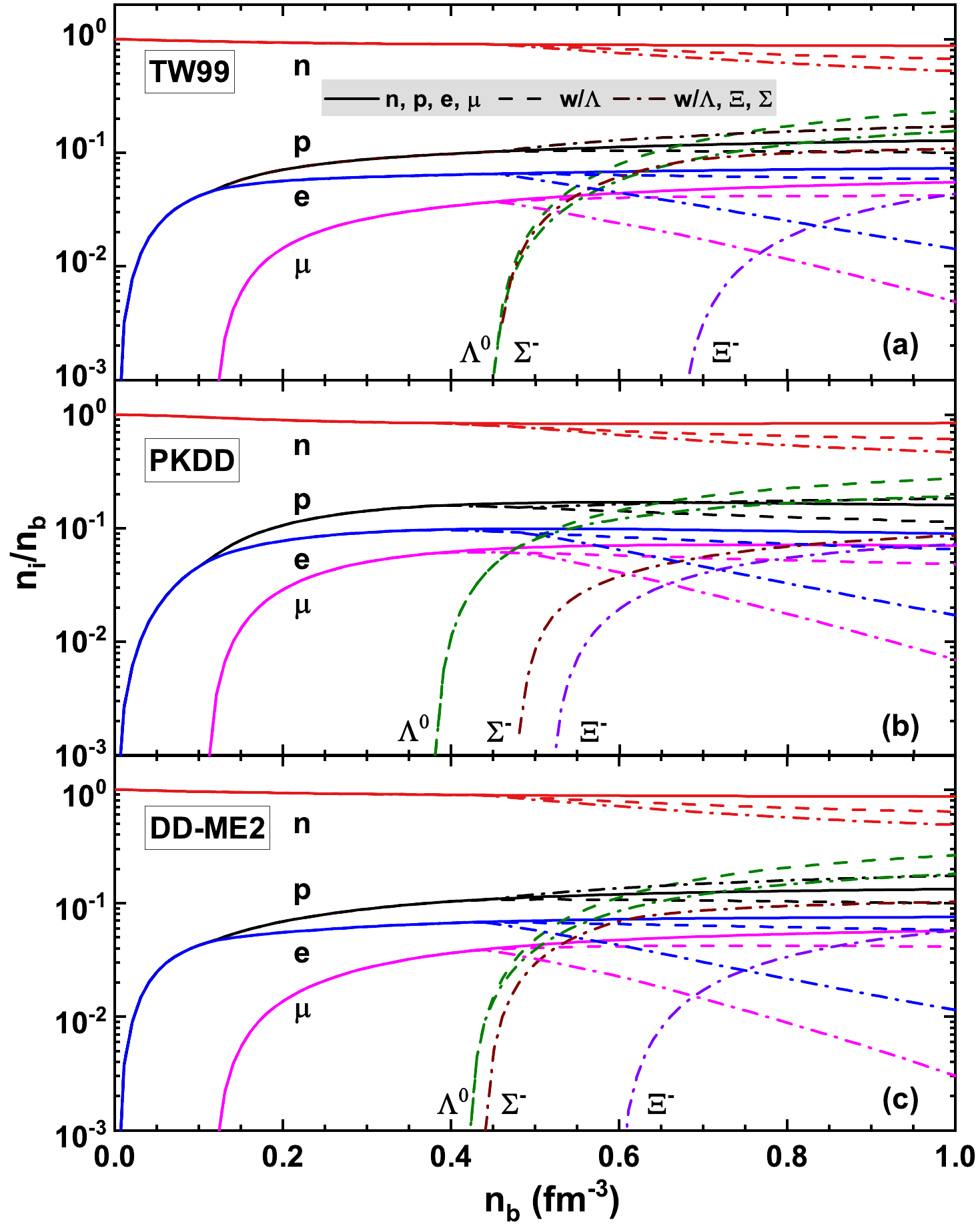}
	\caption{Particle fractions $y_{i}=n_{i}/n_{b}$ for baryons ($n$, $p$, $\Lambda$, $\Xi$, $\Sigma$) and leptons ($e$, $\mu$) in neutron star matter as a function of the total baryon number density $n_{\rm b}=n_{\rm b}^{\rm B}$. Results are shown for three compositional models: nucleons and leptons only (solid lines), with the addition of $\Lambda$ hyperons (dashed lines), and with $\Lambda$, $\Xi$, and $\Sigma$ hyperons (dash-dotted lines). Panels (a), (b), and (c) correspond to calculations based on the TW99~\cite{NPA1999Typel_656_331}, PKDD~\cite{PRC2004Long_69_034319}, and DD-ME2~\cite{PRC2005Lalazissis_71_024312} density functionals, respectively.}
	\label{Fig1Fraction}
\end{figure}

\section{Results and Discussion}
\label{Sec:Results}

FIG.~\ref{Fig1Fraction} shows the particle fractions $y_{i}=n_{i}/n_{b}$ of key baryons and leptons in neutron star matter as a function of the total baryon number density $n_{\rm b}$. The results are compared in three cases with different particle compositions: (i) nucleons and leptons only (solid lines), (ii) inclusion of $\Lambda$ hyperons (dashed lines), and (iii) inclusion of $\Lambda$, $\Xi$, and $\Sigma$ hyperons (dash-dotted lines). Calculations are performed using three density functionals: TW99~\cite{NPA1999Typel_656_331}, PKDD~\cite{PRC2004Long_69_034319}, and DD-ME2~\cite{PRC2005Lalazissis_71_024312}, presented in panels (a), (b), and (c), respectively.

In all compositional scenarios, neutrons remain the dominant baryonic component throughout the displayed density range. Protons, electrons, and muons collectively maintain charge neutrality, though their particle fractions remain significantly lower than that of neutrons by several orders of magnitude. Upon the onset of hyperons, the particle fractions $n_{i}/n_{b}$ undergo substantial changes due to the additional degrees of freedom and the requirements of chemical equilibrium and charge conservation. The $\Lambda$ hyperon is the first strange baryon to emerge, typically at $n_{\rm b} \approx 0.4\,{\rm fm}^{-3}$, depending on the adopted functional. In the case of only including $\Lambda$ hyperons (dashed curves), the particle fractions for nucleons and leptons are all reduced noticeably. Further inclusion of $\Xi$ and $\Sigma$ hyperons (dash-dotted curves) leads to a more diverse hyperon population at higher densities, further suppressing the neutron particle fractions while slightly enhancing the proton fraction. Since the $\Xi^-$ and $\Sigma^-$ hyperons are negatively charged, their presence strongly diminishes the need for electrons and muons to balance the positive charge of protons. It is noted that $\Xi^0$ and $\Sigma^{0,+}$ hyperons do not appear in the displayed density range.

In all three panels of Fig.~\ref{Fig1Fraction}, while the qualitative trends in particle fractions are consistent across the TW99, PKDD, and DD-ME2 density functionals, clear quantitative differences emerge, particularly in the threshold densities for hyperon appearance and in the relative abundances of different hyperon species. These differences can be traced to the distinct saturation properties predicted by PKDD and DD-ME2, as summarized in Table~\ref{Table2}. Specifically, PKDD yields a larger symmetry energy (reflected in higher values of $J$ and $L$), whereas DD-ME2 produces stiffer symmetric nuclear matter (characterized by larger values of $K$ and $Q$). This contrast in the behavior of symmetric and asymmetric nuclear matter likely leads to differing predictions for the onset of new particle degrees of freedom at high baryon number density. Overall, these discrepancies underscore the model dependence in predicting the detailed composition of dense hadronic matter, which directly influences the equation of state.

%------------------------------Fig 2---------------------------------------- 
\begin{figure}[t!]
  \centering
  \includegraphics[width=0.95\linewidth]{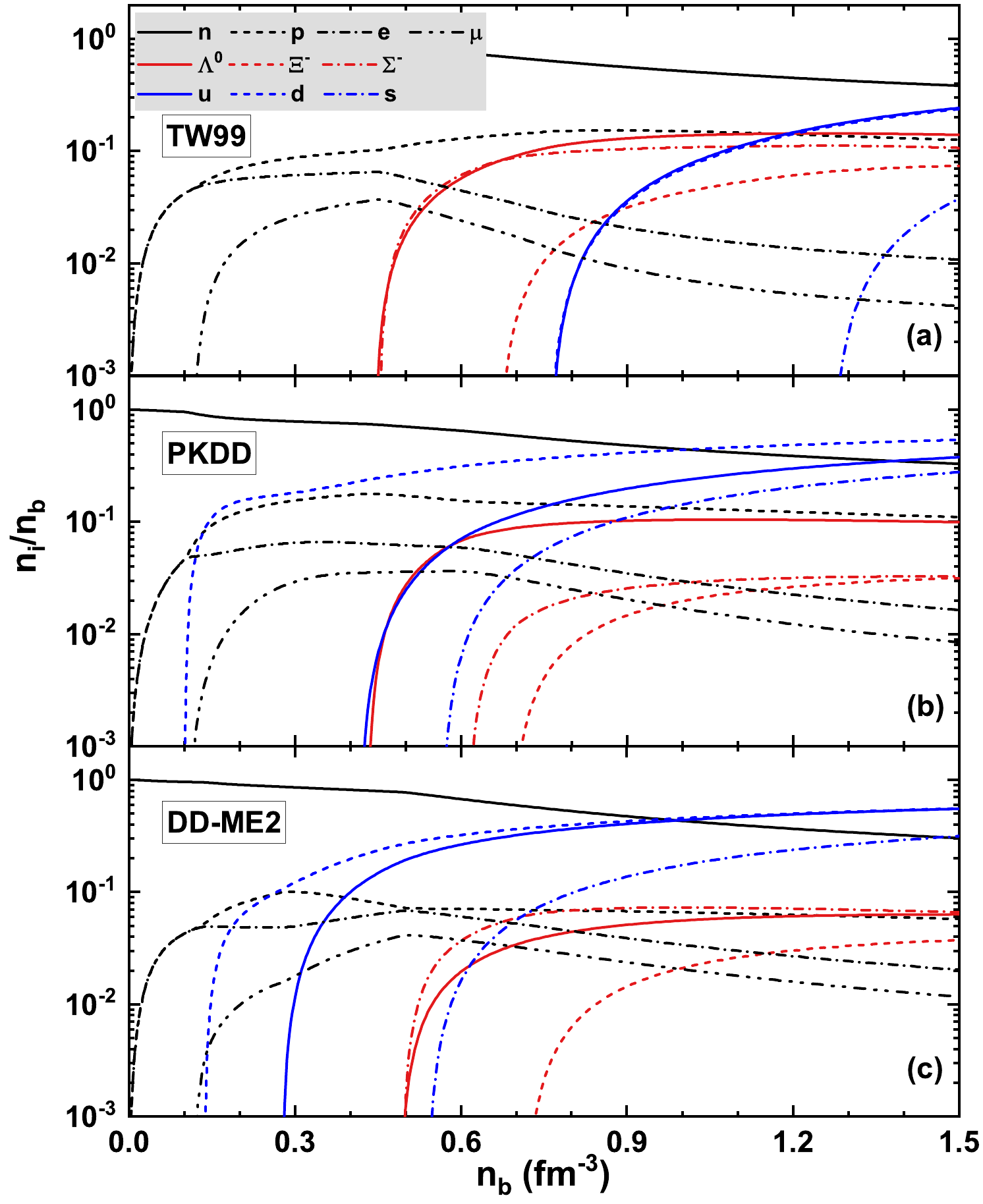}
  \caption{Particle fractions $y_{i}=n_{i}/n_{b}$ for baryons~($n$, $p$, $\Lambda$, $\Xi$, $\Sigma$), quarks~($u$, $d$, $s$), and leptons~($e$, $\mu$) in quarkyonic matter as a function of the total baryon number density $n_{\rm b}=n_{\rm b}^{\rm B}+n_{\rm b}^{\rm Q}$. The parameter sets ($B$, $C$, $\sqrt{D}$) employed based on TW99~\cite{NPA1999Typel_656_331}, PKDD~\cite{PRC2004Long_69_034319}, and DD-ME2~\cite{PRC2005Lalazissis_71_024312} density functionals are (300, 0.7, 180), (150, 0.7, 150), and (100, 0.5, 160), respectively.}
  \label{Fig2Fraction2}
\end{figure}
  
In Fig.~\ref{Fig2Fraction2}, we extend the purely hadronic matter presented in Fig.~\ref{Fig1Fraction} to quarkyonic matter by incorporating quark degrees of freedom ($u, d, s$) and show the particle fractions $y_{i}=n_{i}/n_{b}$ of baryons, leptons, and quarks as a function of the total baryon number density $n_{\rm b}=n_{\rm b}^{\rm B}+n_{\rm b}^{\rm Q}$. Calculations are done based on three density functionals TW99, PKDD, and DD-ME2 for hadronic matter, with corresponding parameter sets $(B, C, \sqrt{D})$ for the baryon and quark mass scalings taken as (300, 0.7, 180), (150, 0.7, 150), and (100, 0.5, 160) respectively when go further to quarkyonic matter. Previous studies in Ref.~\cite{PRD2023Xia_108_054013} have demonstrated that the predicted radii $R_{1.4}$ and tidal deformability $\Lambda_{1.4}$ of 1.4-solar-mass compact stars, as well as the maximum mass $M_{\rm TOV}$ and the maximum equilibrium sound velocity $v_{eq}^{\rm max}$ of neutron star based on these parameters are in better agreement with astrophysical observational constraints. The key feature is the hadron-to-quark transition occurring at a transition baryon number density $n_{\rm b}^{\rm tr}$. 

As shown in panel (a), below this density $n_{\rm b}^{\rm tr}$, the composition closely resembles that of the hadronic matter in Fig.~\ref{Fig1Fraction}, where baryons ($n, p, \Lambda, \Xi, \Sigma$) and leptons dominate. Above $n_{\rm b}^{\rm tr}$, quark degrees of freedom become accessible and the particle fractions of the $d$, $u$, and eventually $s$ quarks increase quickly. In panel (b), it is noted that the $n_{\rm b}^{\rm tr}$ obtained with the PKDD-based parameter set is very low, as low as $n_{\rm b}^{\rm tr} \approx 0.1$~fm$^{-3}$, where the $d$ and $u$ quarks emerge before hyperons at very low baryon number densities, indicating the onset of quarkyonic matter. In this case, a rapid rearrangement of particle components takes place above the transition density $n_{\rm b}^{\rm tr}$: the particle fractions of baryons decrease slightly or keep stable with $n_{\rm b}$ while the $d, u, s$ quarks increase drastically and become the dominant components of the system in high baryon number density. The dominance of the $u$ and $d$ quarks, together with various hyperons ($\Lambda$, $\Xi^{-}$, $\Sigma^{-}$) in the high-density regime indicates that the structure of high-density neutron star must incorporate contributions from both quark and hyperon degrees of freedom. These conclusions are also shown in panel (c) for results obtained with the DD-ME2-based parameter set except a slightly higher transition density $n_{\rm b}^{\rm tr}\approx 0.14$~fm$^{-3}$. This reflects a shift in the effective degrees of freedom from confined baryons to deconfined quarks within the quarkyonic framework. 

%------------------------------Fig 3---------------------------------------- 
\begin{figure}[t!]
	\centering
	\includegraphics[width=0.95\linewidth]{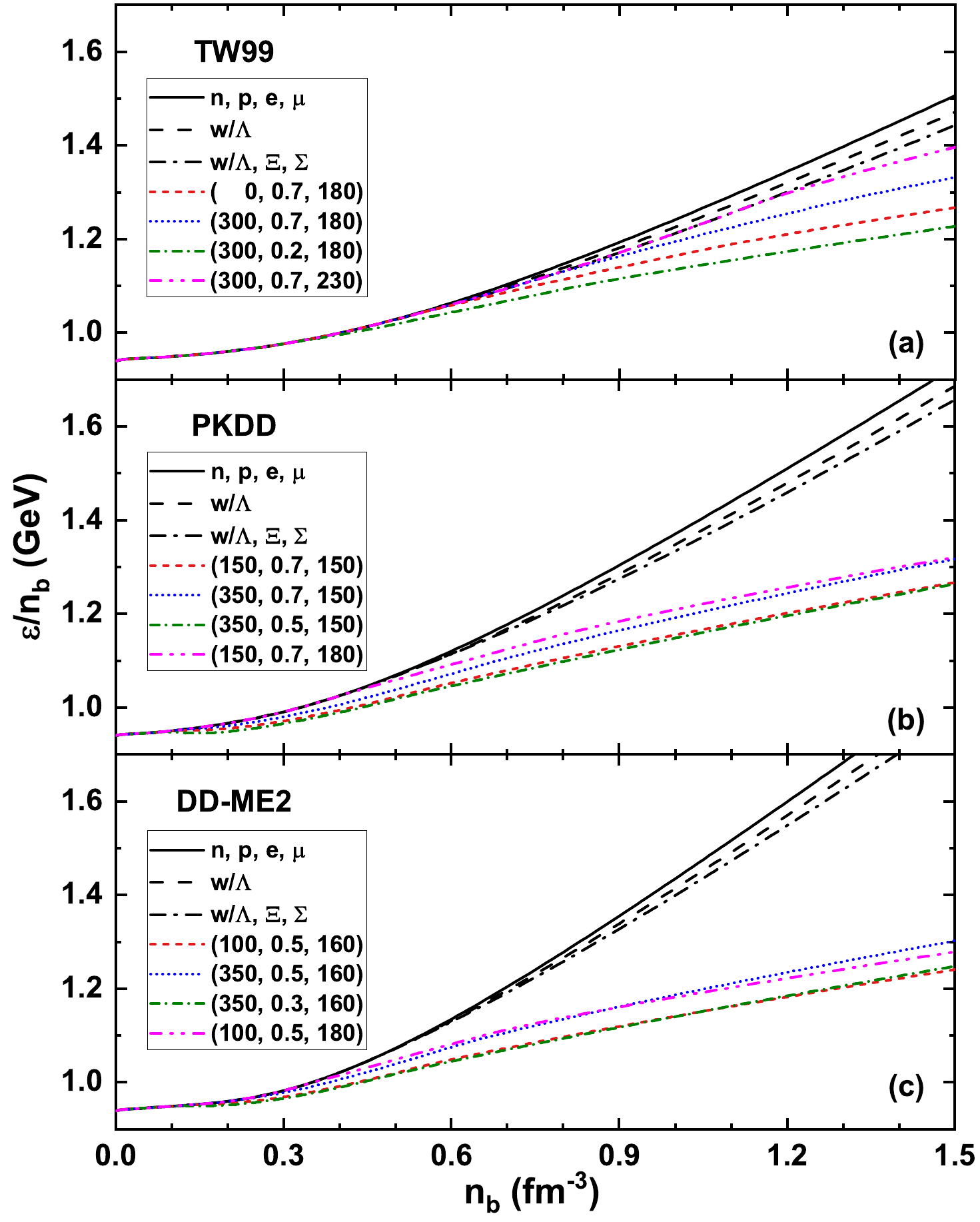}
    \caption{Energy per baryon $\varepsilon/n_{\rm b}$ as a function of the total baryon number density $n_{\rm b}$ for nuclear matter (black lines) composted with nucleons and leptons only (solid lines), the further addition of $\Lambda$ hyperons (dashed lines), and the full inclusion of $\Lambda$, $\Xi$, and $\Sigma$ hyperons (dash-dotted lines), and quarkyonic matter (colored lines). Panels~(a), (b), and (c) correspond to calculations based on the TW99~\cite{NPA1999Typel_656_331}, PKDD~\cite{PRC2004Long_69_034319}, and DD-ME2~\cite{PRC2005Lalazissis_71_024312} density functionals, respectively, together with the $(B, C, \sqrt{D})$ parameter sets listed in Table~\ref{Table3}.}
	\label{Fig3Enb}
\end{figure}

In Fig.~\ref{Fig3Enb}, we present the energy per baryon $\varepsilon/n_{\rm b}$ as a function of the total baryon number density $n_{\rm b}$, providing a direct comparison between purely hadronic matter and quarkyonic matter. The purely hadronic matter (black lines) with different particle compositions show a systematic trend: the equation of state (EOS) softens with the additions of hyperons. The nucleon-only EOS (solid line) has the highest energy at high densities while the inclusion of $\Lambda$ hyperons (dashed line) reduces $\varepsilon/n_{\rm b}$, and the further addition of $\Xi$ and $\Sigma$ hyperons (dash-dotted line) reduces it even more. This illustrates the well-known hyperon softening effect, where new degrees of freedom reduce the pressure at a given density, making the EOS less stiff. The quarkyonic matter (colored lines) introduce a more dramatic change for $\varepsilon/n_{\rm b}$. In general, the incorporation of quark degrees of freedom ($u$, $d$, $s$) contributes to additional EOSs softening. At the transition baryon number density $n_{\rm b}^{\rm tr}$, the curves for quarkyonic matter deviates from those for the hadronic matter, the positions of which are also predicted by the appearance of $d$ quarks in Fig.~\ref{Fig2Fraction2}. For example, based on the PKDD parameter set shown in panel (b), the transition from the hadronic to the quarkyonic phase occurs at $n_{\rm b}^{\rm tr} \gtrsim 0.1$~fm$^{-3}$, accompanied by a decrease in energy, which indicates enhanced thermodynamic stability at higher baryon densities. Furthermore, the ($B, C, \sqrt{D}$) parameters have great impacts, where increasing the quark-hadron interaction strength $B$, perturbative interaction $C$, and confinement $\sqrt{D}$ all results in higher onset densities for the quarkyonic transition and energies of quarkyonic matter. Notably, an increase in $C$ leads to a particularly pronounced rise in energy at high densities. These findings are consistent with our previous studies~\cite{PRD2023Xia_108_054013}.

%------------------------------Fig 4---------------------------------------- 
\begin{figure}[t!]
	\centering
	\includegraphics[width=0.95\linewidth]{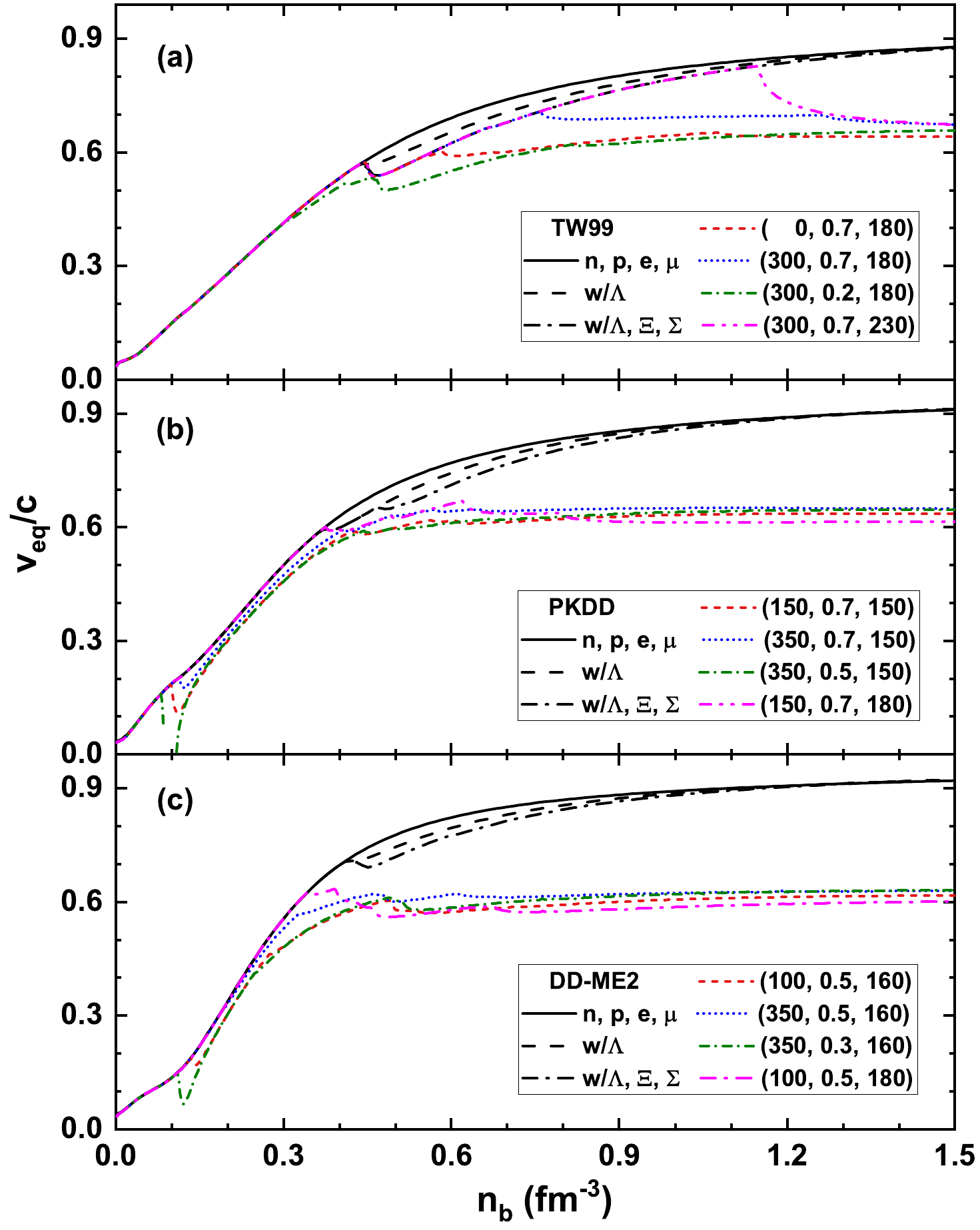}
	\caption{The same as Fig.~\ref{Fig3Enb}, but for the equilibrium sound velocity $v_{eq}/c$.}
	\label{Fig4velocity}
\end{figure}

To more explicitly illustrate the variations in the stiffness of the EOSs, we also analyze the equilibrium sound velocity $v_{eq}/c$, defined as the total derivative of the pressure with respect to energy density,  
\begin{equation} 
v_{eq}/c=\sqrt{\displaystyle\frac{dP}{d\varepsilon}},
\end{equation}
which is a key observable for investigating phase transition. This quantity is related to the ratio of the compressibility to the total baryon chemical potential and thus serves as a measure of the elastic properties of a compact star. It is noted that the equilibrium sound velocity $v_{eq}/c$ is different from the adiabatic sound speed $v_{ad}/c=\sqrt{(\partial P/\partial\varepsilon)|_{y_{lep}}}$. The $v_{ad}/c$ is relevant in hydrodynamical contexts such as simulations of core-collapse supernovae and of binary mergers of compact stars. The difference between these two sound speeds leads to $g$-mode oscillations, and several recent studies~\cite{MNRAS2014Kantor_442_90,MNRAS2016Dommes_455_2852,MNRAS2017Yu_464_2622,MNRAS2018Rau_481_4427,PRD2022Zhao_105_103025,PRD2023Zheng_107_103048} have argued that exotic matter in the cores of compact stars can be inferred from the analysis of their $g$-mode spectra. Although there is no experimental measurements of the equilibrium sound velocity in high density, perturbative QCD theory establishes a limit for the equilibrium sound velocity in quark-gluon plasma, given by $v_{eq} = 1/\sqrt{3} \ (\sim 0.58\,c)$~\cite{PRD1997Freedman_16_1169,PRD2005Fraga_71_105014,PRD2010Kurkela_81_105021}. Given that the precise baryon number density at which the hadron-quark phase transition occurs is still uncertain, a measured equilibrium sound velocity could shed light on the phase transition.

Figure~\ref{Fig4velocity} displays the equilibrium sound velocity $v_{eq}/c$ as a function of baryon number density $n_{\rm b}$, providing a complementary perspective on the stiffness of the equation of state following the energy analysis presented in Fig.~\ref{Fig3Enb}. For the purely hadronic matter (black lines), the results quantify the progressive softening induced by hyperons: the nucleon-only EOS (solid line) exhibits the highest equilibrium sound velocity, while the sequential addition of $\Lambda$ (dashed) and then all $\Xi$ and $\Sigma$ hyperons (dash-dotted) systematically lowers $v_{eq}/c$ across wide density range. This confirms that hyperon degrees of freedom reduce the pressure. However, at high baryon number densities, the sound speeds $v_{eq}/c$ of hadronic matter containing different nucleons/hyperons tend to be consistent, and the results calculated with parameter sets TW99, PKDD, and DD-ME2 are $0.89$, $0.91$, and $0.92$ respectively, which are much higher than the value of $0.58\,c$. In the case of quarkyonic matter (colored lines), a pronounced drop in $v_{eq}/c$ is observed around the transition baryon number density $n_{\rm b}^{\rm tr}$ where quark degrees of freedom appear. Above $n_{\rm b}^{\rm tr}$, the equilibrium sound velocity in the quarkyonic phase is significantly lower than that in nuclear matter, with its maximum value $v_{eq}^{\rm max}$ can be reduced to as low as $0.6\,c$, much closer to the ultrarelativistic limit $0.58\,c$. 

Moreover, near the density where new degrees of freedom begin to emerge, the equilibrium sound velocity $v_{eq}/c$ exhibits a nonmonotonic trend, reflecting its sensitivity to compositional changes. Taking panel (a) of Fig~\ref{Fig4velocity} as an example, for pure hadronic matter, when the total baryon number density reaches approximately $0.4\,\text{fm}^{-3}$, the equilibrium sound velocity shows a nonmonotonic behavior characterized by an initial decrease followed by an increase, due to the appearance of hyperon degrees of freedom. For quarkyonic matter, the nonmonotonic behavior of the equilibrium sound velocity varies depending on the parameter sets $(B,C,\sqrt{D})$. Additionally, similar features can also be observed in panels (b) and (c). Furthermore, the maximum equilibrium sound velocity $v_{eq}^{\rm max}$ depends on the choice of $(B, C, \sqrt{D})$ parameter sets, with smaller values of $B$, $C$, and $\sqrt{D}$ leading to lower $v_{eq}^{\rm max}$, as illustrated in Table~\ref{Table3} and Fig.~\ref{Fig4velocity}. Notably, a first order phase transition exhibits when employing ($B=350$, $C=0.5$, $\sqrt{D}=150$~MeV) based on the PKDD parameter set, where the equilibrium sound velocity vanishes in the baryon density range $n_{\rm b} \approx 0.087\sim 0.1$~fm$^{-3}$.

%------------------------------Fig 5---------------------------------------- 
\begin{figure*}[t!]
\centering
\includegraphics[width=0.95\linewidth]{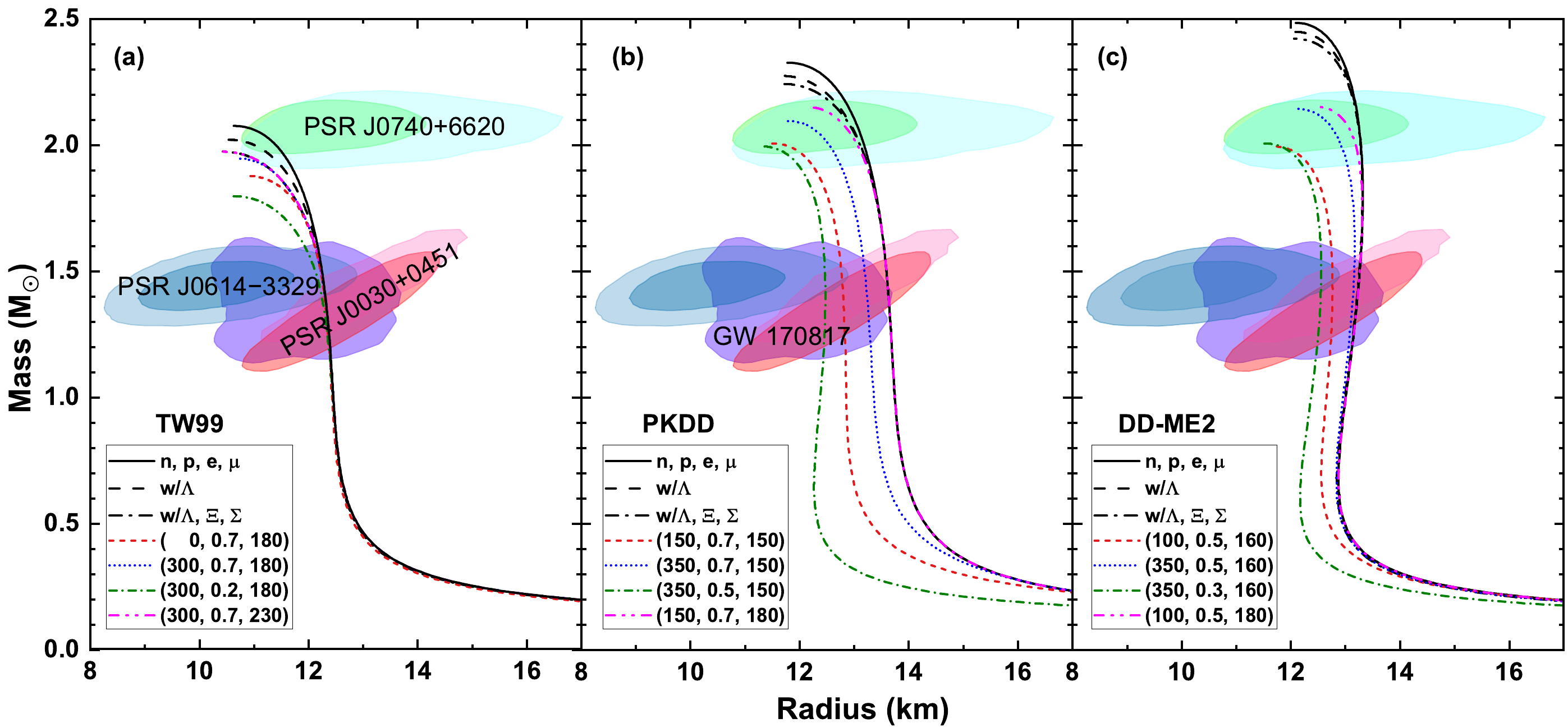}
\caption{Mass-radius relations of compact stars derived EOS presented in Fig.~\ref{Fig3Enb}. The shaded regions represent the constraints from the binary neutron star merger event GW170817~($90\%$ credible region)~\cite{PRL2018Abbott_121_161101}, as well as the observational pulse profiles of PSR J0030+0451~($68\%$ credible region)~\cite{AJL2019Riley_887_L21,AJL2019Miller_887_L24}, PSR J0740+6620~($68\%$ credible region)~\cite{AJL2021Riley_918_L27,AJL2021Miller_918_L28}, and PSR J0614-3329~(inner and outer regions corresponding to the 68\% and 95\% credible regions, respectively)~\cite{ApJ2025Mauviard_995_60}. Panels~(a), (b), and (c) correspond to calculations based on the TW99~\cite{NPA1999Typel_656_331}, PKDD~\cite{PRC2004Long_69_034319}, and DD-ME2~\cite{PRC2005Lalazissis_71_024312} density functionals, respectively, together with the $(B, C, \sqrt{D})$ parameter sets listed in Table~\ref{Table3}.}
\label{Fig5MR}
\end{figure*}

Based on the EOSs shown in Fig.~\ref{Fig3Enb}, we determine the structure of compact stars by solving the TOV equation~(\ref{TOV}). Figure~\ref{Fig5MR} presents the mass-radius ($M$-$R$) relations for compact stars which are compared with the astrophysical observational constrains, i.e., the $90\%$ credible region from the binary neutron star merger GW170817 ($M=1.36_{-0.18}^{+0.22} M_{\odot}$, $R_{1}=11.9 \pm 1.4$~km)~\cite{PRL2018Abbott_121_161101}, and the $68\%$ credible regions from the observational pulse profiles of PSR J0030+0451 (${\displaystyle M=1.34 ^{+0.15}_{-0.16} M_{\odot}}$, ${\displaystyle R=12.71^{+1.14}_{-1.19}}$~km)~\cite{AJL2019Riley_887_L21,AJL2019Miller_887_L24}, PSR J0740+6620 ($M$$=$$2.08 \pm 0.07 M_{\odot}$, ${\displaystyle R=13.7^{+2.6}_{-1.5}}$~km)~\cite{AJL2021Riley_918_L27,AJL2021Miller_918_L28}, and PSR J0614-3329 (${\displaystyle M=1.44^{+0.06}_{-0.07} M_{\odot}}$, ${\displaystyle R=10.29^{+1.01}_{-0.86}}$~km)~\cite{ApJ2025Mauviard_995_60}. 

In the hadronic phase~(black lines), the predictions vary significantly with the chosen density functional. Based on the TW99 parameter set, the mass and radius of compact stars containing only nucleons (solid line) reproduce the observational constraints well, with the maximum mass around $2.1 M_{\odot}$. However, when hyperons are included, first only the $\Lambda$ hyperons (dashed line), then the $\Lambda$, $\Xi$, and $\Sigma$ hyperons together (dash-dotted line), the EOSs become softer as shown in Fig.~\ref{Fig3Enb}, leading to a gradual reduction in the predicted maximum mass $M_{\rm TOV}$. In the case with all hyperons included, $M_{\rm TOV}$ falls outside the observationally allowed region, indicating that such a soft EOS is inconsistent with current astrophysical data. In contrast, the PKDD and DD-ME2 density functionals produce much stiffer EOSs for compact stars in the hadronic phase. Their $M$-$R$ curves lie above the credible regions from PSR J0740+6620, and $M_{\rm TOV}$ substantially exceeds the mass upper limits, suggesting that pure hadronic matter descriptions with these effective interactions are too stiff to be compatible with neutron star observations due to large $L$~(PKDD) and $Q$~(DD-ME2). It is noted that the hyperon degrees of freedom have little influence on the predicted radius for all three density functionals.

For the quarkyonic phase~(colored lines), the introduction of quark degrees of freedom significantly softens the EOSs for all density functionals. Based on the TW99 parameter set, the transition from hadronic to quarkyonic phase leads to a maximum compact star mass below $2M_{\odot}$, which is seriously deviated from the established observations of large-mass pulsars. In contrast, based on the PKDD and DD-ME2 parameter sets, the pronounced softening effect by the inclusion of quark degrees of freedom shift the $M$-$R$ curves to the credible regions from PSR J0740+6620 rightly, producing more compact hybrid stars with reduced radii $R$ and lower maximum masses $M_{\rm TOV}$. This suggests that the PKDD functional with a larger symmetry energy slope $L$ or DD-ME2 with a larger skewness coefficient $Q$ are more conducive to the onset of a quarkyonic phase transition. Compared to the results reported in Ref.~\cite{PRD2023Xia_108_054013}, the EOSs of quarkyonic matter discussed in this work exhibit stronger softening at high densities, further lowering $M_{\rm TOV}$ and yielding improved consistency with pulsar mass measurements. Furthermore, the stiffness of the quarkyonic $M$-$R$ curves is sensitive to the parameter sets $(B, C, \sqrt{D})$ of mass scalings. Generally, smaller values of those parameters result in a softer EOS, shifting the curves toward smaller radii and lower maximum masses. All those behaviors highlight how the quarkyonic model can improve the description of stiff hadronic matter, leading to consistency with astrophysical data for appropriately chosen parameters.

\section[short]{Summary}
\label{Sec:Summary}

Quarkyonic matter phase plays as a key mechanism for understanding the transition from hadronic matter to quark matter in compact stars.
In this work, we describe the properties of quarkyonic matter by constructing a “quark Fermi sea” with a “baryon Fermi surface” within the framework of the RMF model and the equivparticle model with density-dependent quark masses, which can give a coherent investigation of nuclear matter, quark matter, and quarkyonic matter in a unified manner. As an extension of our previous work~\cite{PRD2023Xia_108_054013}, we further introduce strangeness degrees of freedom, incorporating $\Lambda$, $\Xi$, and $\Sigma$ hyperons together with strange quarks in a unified quarkyonic framework. The present theoretical framework is more comprehensive, offering a more complete and reliable description of the multiphase behavior of strongly interacting matter under high-density conditions. The interactions between baryons are described by the exchange of $\sigma$, $\omega$, and $\rho$ mesons while quarks are treated as quasifree particles with density-dependent masses, effectively incorporating confinement and leading-order perturbative interactions. Three density functional models, TW99~\cite{NPA1999Typel_656_331}, PKDD~\cite{PRC2004Long_69_034319}, and DD-ME2~\cite{PRC2005Lalazissis_71_024312}, are employed in the analysis, each exhibiting distinct saturation properties of nuclear matter. Specifically, PKDD yields a larger symmetry energy slope $L$, DD-ME2 features a higher skewness coefficient $Q$, and TW99 demonstrates favorable agreement with empirical constraints derived from neutron star observations and heavy-ion collision experiments. This work presents a comprehensive study of quarkyonic phase with strangeness degrees of freedom in compact stars. 

Our calculations show that the inclusion of hyperons softens the EOSs, reducing the equilibrium sound velocity around $n_{\rm b} \approx 2n_0$ and consequently lowering the predicted masses and radii of neutron stars. Furthermore, when a quark–hadron phase transition is incorporated, the EOSs undergo additional softening at high densities. In this case, the maximum equilibrium sound velocity reaches $v_{eq}^{\rm max} \approx 0.6\,c$, which is very close to the ultrarelativistic limit of $0.58\,c$. The maximum Tolman–Oppenheimer–Volkoff mass $M_{\rm TOV}$ is also significantly reduced to $2.0~M_{\odot}$, which is in good agreement with current astrophysical constraints~\cite{AJL2019Riley_887_L21,AJL2021Riley_918_L27,AJL2019Miller_887_L24,AJL2021Miller_918_L28}. These trends are particularly evident when using the PKDD functional with a larger symmetry energy slope $L$, or the DD-ME2 parameterization with a larger skewness coefficient $Q$.

\begin{acknowledgments}
This work was supported by the Natural Science Foundation of Henan Province (Grant No.~242300421156), the National Natural Science Foundation of China (Grants No.~U2032141, No.~12275234, and No.~12205057), and the National SKA Program of China (Grant No.~2020SKA0120300).		
\end{acknowledgments}

\section*{Data availability}		
The data that support the findings of this article are openly available~\cite{data}, embargo periods may apply.

%\bibliography{bibliography}

\end{document}